# Decision Structure of Risky Choice

Lamb Wubin[*]   Naixin Ren

January 30, 2017

As we know, there is a controversy about the decision making under risk between economists and psychologists. We discuss to build a unified theory of risky choice, which would explain both of compensatory and non-compensatory theories. For risky choice, according to cognition ability, we argue that people could not build a continuous and accurate subjective probability world, but several order concepts, such as small, middle and large probability. People make decisions based on information, experience, imagination and other things. All of these things are so huge that people have to prepare some strategies. That is, people have different strategies when facing to different situations. The distributions of these things have different decision structures. More precisely, decision making is a process of simplifying the decision structure. However, the process of decision structure simplifying is not stuck in a rut, but through different path when facing problems repeatedly. It is why preference reversal always happens when making decisions. The most efficient way to simplify the decision structure is calculating expected value or making decision based on one or two dimensions. We also argue that the deliberation time at least has four parts, which are consist of substitution time, $\tau''$ $(G)$ $d\tau$ time, $\tau'$ $(G)$ $d\tau$ time and calculation time. Decision structure also can simply explain the phenomenon of paradoxes and anomalies. JEL Codes: C10, D03, D81

## I. INTRODUCTION

What is the origin of risk? It is possible to construct very special cases. My nephew, little Rugmann, is 6 months old. One day he got a mystery box from Schrödinger. His greatest pleasure is opening this box before going to sleep. Imagine little Rugmann play with the Schrödinger's cat every night. However, the morning sun never lasts a day. Several days later, he opens the box and finds a dead cat. At this time, little Rugmann is aware of risk, a risky box, a risky cat. He is so worried that he has not enough courage to open this box every night. In other words, the meaning of origin of risk can be illustrated by an idiom that they who know nothing fear nothing.

What is the human decision making? Sometimes we have to acknowledge that the experimental studies are different from the real life decisions. Decision strategy is not stuck in a rut, but based on the things, in the real life, and experiment materials, in the laboratory. So we could not build a theory without discussing them. It is no doubt that, according to our cognition ability, we could not build a continuous and accurate subjective probability world. That is, we can only establish several intervals of probability perception.

[*]Institute of Psychology, Chinese Academy of Sciences; University of Chinese Academy of Sciences; 16th, Lincui Road, Chaoyang District, Beijing, P. R. China 100101. I'd like to thank everyone who gave comments and suggestions. I'm looking forward to discussing, such as mathematics, decision making, and grammar with you. Email: wubin1991122@126.com.

As we know, there is a controversy about the decision making under risk between economists and psychologists. For the most of economists, risky choice is based on a compensatory expectation-maximizing process. Alternatively, the most of psychologists think that decision making is a trading off process between different dimensions, such as risk dimension, value dimension. Necessarily, however, we strictly believe that we have to build a unified theory of risky choice, which would explain both of compensatory and non-compensatory theories. Note that, since gain and loss are two different parts of the decision under risk. For simplicity, we will just discuss the gain of risky choice and neglect the influence of non-strategy, such as emotion, intelligence, gender and others.

From a metaphysical point of view, first of all, we will discuss origin of risk, namely that the perception of probability or risk. It is no doubt that risk and probability are abstract concepts, which were acquired subsequently. With increasing age, little Rugmann establishes two constant concepts of probability, and the first cognition of risky or probability is 0, which means gain nothing. Then, he is promptly recognized the meaning of 1, that is gain everything. After that, Rugmann at least develops three concepts, near to zero $a$, near to one $c$, between zero to one $b$. All of five concepts constitute the first rudiment of probability system, which is called probability set. Moreover, with development of experience, new probability, p = 0.1, will be substituted for an appropriate element, near to zero $a$, of the probability set.

## II. DECISION STRUCTURE

Decision making under risk is a process of simplifying the decision structure, which is distribution of probabilities and rewards in the decision matrix. Let us give a new mathematical expression of decision making, which is called decision matrix $G$. More precisely, a relative position relationship of elements can be defined as a decision matrix. The top left corner of decision matrix shows the smallest element, and the bottom left corner shows the largest element. For instance, $G_1$ was showed as follow,

$$G_1 = \begin{pmatrix} i & j \\ 1 & k \end{pmatrix}, i< j< k<1. \quad (1)$$

We first introduce some basic definitions for the sake of completeness.

Definition 1. Let us define the operation of decision matrix, which is the Hadamard product of $G_1$ and $G_2$, denoted by $G_1 \circ G_2$, the decision matrix as follows.

$$G = G_1 \circ G_2 = \begin{pmatrix} i & j \\ l & k \end{pmatrix} \circ \begin{pmatrix} i' & j' \\ l' & j' \end{pmatrix}$$

$$= \begin{pmatrix} ii' & jj' \\ ll' & kk' \end{pmatrix} \quad 0< aa'< bb'< cc'< ll'. \quad (2)$$

Definition 2. Let $S_p$ be the set of all probabilities. Now define the set of all acquired probabilities.

$$G_p = \{a \in S_p \mid a \text{ was acquired subsequently }\}$$

Definition 3. The set of all substitution functions from $S_p$ to $G_p$ is denoted by {f:

$S_p \to G_p$}.

$$S_p (0) \to G_p (0);$$
$$S_p (P_1, P_2 \ldots P_{k-1}) \to G_p (a);$$
$$S_p (P_k, P_{k+1} \ldots P_{m-1}) \to G_p (b);$$
$$S_p (P_m, P_{m+1} \ldots P_n) \to G_p (c);$$
$$S_p (1) \to G_p (1)$$

Thus,
$$G_p = \{0, a, b, c, 1\}$$

For simplicity, the decision matrix of probability $G_p$ as follows,

$$G_p = \begin{pmatrix} a & b \\ 1 & c \end{pmatrix}, \text{ a< b< c<1.} \quad (3)$$

In the same way, we also give the expression of rewards, money, in the risky choice. The first rudiment of value system has two constant elements: zero and infinity, three concepts: $a'$ = small, $b'$ = mean, $c'$ = large.

Definition 4. Let $S_V$ be a set of rewards, $S_V \in (0, \infty)$. $G_V = \{0, a', b', c', \infty\}$ is a subset of $S_V$. Decision matrix of value as follow,

$$G_V = \begin{pmatrix} a' & b' \\ \infty & c' \end{pmatrix}, a' < b' < c' < \infty. \quad (4)$$

Assumptions 1. The substitution does not stop until human can compare a superiority option, which must be substituted for a superiority position in the decision matrix.

Assumptions 2. A new probability or value will be substituted for appropriate element in the decision matrix.

It is assumed that there are two options: A (Pi, Vi) vs. B (Pj, Vj), Pi, Pj $\in S_p$, Vi, Vj $\in S_v$. For instance,

1) $G_a = G_p \circ G_v = \begin{pmatrix} a_i a_i' & b_j b_j' \\ 1 & 1 \end{pmatrix}$, the option B always has the superiority position in the decision matrix. This situation was called Zero Order Substitution, $\tau (G)$.

2) $G_b = G_p \circ G_v = \begin{pmatrix} a_i a_j a_i' & b_j' \\ 1 & 1 \end{pmatrix}$, the option B not always has the superiority position in the decision matrix. This situation was called First Order Substitution, $\tau' (G)$.

3) $G_c = G_p \circ G_v = \begin{pmatrix} a_i a_j' & b_j b_i' \\ 1 & 1 \end{pmatrix}$, we don't know, options crossing, which option has the superiority position in the decision matrix. This situation was called Second Order Substitution, $\tau'' (G)$.

According to Assumptions 1 and $G_c$, we can deduce that Second Order Substitution will simplify to First Order Substitution during decision making process, which was called order reduction, denoted by $d\tau$.

Thus,

4) $G_d = \tau'' (G_c) d\tau = \begin{pmatrix} a_j' & b_i b_j b_i' \\ 1 & 1 \end{pmatrix}$, this situation was called risk seeking.

5) $G_e = \tau'' (G_c) d\tau = \begin{pmatrix} a_i a_j a_j' & b_i' \\ 1 & 1 \end{pmatrix}$, this situation was called risk aversion.

6) $G_d = \tau''(G_c)\, d\tau = \begin{pmatrix} a_i & b_j b_i' b_j' \\ 1 & 1 \end{pmatrix}$, this situation was called value seeking.

7) $G_e = \tau''(G_c)\, d\tau = \begin{pmatrix} a_i a_i' a_j' & b_j \\ 1 & 1 \end{pmatrix}$, this situation was called value aversion.

What is decision structure? Imagine that there are two cups. Which one do you prefer? The cups have a number of attributes such as color, size and weight. The distribution of attribute is called decision structure. If all attribute of cup A is better than cup B, decision structure theory predicts that people will choose cup A. This structure is called zero order substitution. That is, we need not substitution anymore. So, what is substitution? If I tell you the win probability of gamble is 0.1, you will think, wow, how small this probability is. Actually, 0.1 is number, and small probability is concept. The first step, 0.1 is substituted for small probability. This is a substitution. If some attributes are closely, this structure is called first order substitution. Namely that people have to substitute this structure for zero order or make decision based on one or two attributes. If all attributes are cross, this structure is called second order substitution. In this structure, people have to substitute twice or more, or calculate expected value. In a word, decision making is a process of simplifying the decision structure, from complex to easy or familiar. Next step, there are 3 main structures.

**(i). Zero Order Substitution, $\tau(G)$:**

$$G_p = \begin{pmatrix} a_i & b_j \\ 1 & 1 \end{pmatrix} \quad (5)$$

$$G_p = \begin{pmatrix} a_i & 1 \\ 1 & c_j \end{pmatrix} \quad (6)$$

$$G_p = \begin{pmatrix} 1 & b_i \\ 1 & c_j \end{pmatrix} \quad (7)$$

$$G_v = \begin{pmatrix} a_i' & b_j' \\ 1 & 1 \end{pmatrix} \quad (8)$$

$$G_v = \begin{pmatrix} a_i' & 1 \\ 1 & c_j' \end{pmatrix} \quad (9)$$

$$G_v = \begin{pmatrix} 1 & b_i' \\ 1 & c_j' \end{pmatrix} \quad (10)$$

$$(5) \circ (8) = \begin{pmatrix} a_i a_i' & b_j b_j' \\ 1 & 1 \end{pmatrix} \quad (11)$$

In this case, obviously, we can deduce that option B (Pj, Vj) is always appearing at the superiority position of decision matrix. Thus, we do not want to deeply discuss this situation.

**(ii). First Oder Substitution, $\tau'(G)$:**

Section 1

$$G_p = \begin{pmatrix} a_i a_j & 1 \\ 1 & 1 \end{pmatrix} \quad (12)$$

$$G_p = \begin{pmatrix} 1 & b_i b_j \\ 1 & 1 \end{pmatrix} \quad (13)$$

$$G_p = \begin{pmatrix} 1 & 1 \\ 1 & c_i c_j \end{pmatrix} \quad (14)$$

$$G_v = \begin{pmatrix} a_i' & b_j' \\ 1 & 1 \end{pmatrix} \quad (15)$$

$$G_v = \begin{pmatrix} a_i' & 1 \\ 1 & c_j' \end{pmatrix} \quad (16)$$

$$G_v = \begin{pmatrix} 1 & b_i' \\ 1 & c_j' \end{pmatrix} \quad (17)$$

Up to now we considered the structure of decision matrix $G$.

$$(12) \circ (15) = \begin{pmatrix} a_i a_j a_i' & b_j' \\ 1 & 1 \end{pmatrix} \quad (18)$$

$$(12) \circ (16) = \begin{pmatrix} a_i a_j a_i' & 1 \\ 1 & c_j' \end{pmatrix} \quad (19)$$

$$(12) \circ (17) = \begin{pmatrix} a_i a_j & b_i' \\ 1 & c_j' \end{pmatrix} \quad (20)$$

$$(13) \circ (15) = \begin{pmatrix} a_i' & b_i b_j b_j' \\ 1 & 1 \end{pmatrix} \quad (21)$$

$$(13) \circ (16) = \begin{pmatrix} a_i' & b_i b_j \\ 1 & c_j' \end{pmatrix} \quad (22)$$

$$(13) \circ (17) = \begin{pmatrix} 1 & b_i b_j b_i' \\ 1 & c_j' \end{pmatrix} \quad (23)$$

$$(14) \circ (15) = \begin{pmatrix} a_i' & b_j' \\ 1 & c_i c_j \end{pmatrix} \quad (24)$$

$$(14) \circ (16) = \begin{pmatrix} a_i' & 1 \\ 1 & c_i c_j c_j' \end{pmatrix} \quad (25)$$

$$(14) \circ (17) = \begin{pmatrix} 1 & b_i' \\ 1 & c_i c_j c_j' \end{pmatrix} \quad (26)$$

In this section, the probability matrix has same position substitution. However, the value matrix has different size relationship substitution.

Section 2

$$G_p = \begin{pmatrix} a_i & b_j \\ 1 & 1 \end{pmatrix} \quad (27)$$

$$G_p = \begin{pmatrix} a_i & 1 \\ 1 & c_j \end{pmatrix} \quad (28)$$

$$G_p = \begin{pmatrix} 1 & b_i \\ 1 & c_j \end{pmatrix} \quad (29)$$

Up to now we considered the structure of decision matrix $G$.

$$(27) \circ (15) = \begin{pmatrix} a_i a_j' a_i' & b_j \\ 1 & 1 \end{pmatrix} \quad (30)$$

$$(27) \circ (16) = \begin{pmatrix} a_i & b_j b_i' b_j' \\ 1 & 1 \end{pmatrix} \quad (31)$$

$$(27) \circ (17) = \begin{pmatrix} a_i & b_j \\ 1 & c_i' c_j' \end{pmatrix} \quad (32)$$

$$(28) \circ (15) = \begin{pmatrix} a_i a_i' a_j' & 1 \\ 1 & c_j' \end{pmatrix} \quad (33)$$

$$(28) \circ (16) = \begin{pmatrix} a_i & b_i' b_j' \\ 1 & c_j \end{pmatrix} \quad (34)$$

$$(28) \circ (17) = \begin{pmatrix} a_i & 1 \\ 1 & c_j c_i' c_j' \end{pmatrix} \quad (35)$$

$$(29) \circ (15) = \begin{pmatrix} a_i' a_j' & b_i \\ 1 & c_j \end{pmatrix} \quad (36)$$

$$(29) \circ (16) = \begin{pmatrix} 1 & b_i b_i' b_j' \\ 1 & c_j \end{pmatrix} \quad (37)$$

$$(29) \circ (17) = \begin{pmatrix} 1 & b_i \\ 1 & c_j c_i' c_j' \end{pmatrix} \quad (38)$$

In this structure, the most of non-compensatory theories have better fit. Namely, these theories believe that the decision making is trading off between two dimensions, which are probability dimension and reward dimension. Let us considered (18) that we will neglect the difference of probability dimension, and make decision depending on the size difference between Vi and Vj, $\tau(G) = \tau'(G_{18}) d\tau$. In this case, these are the decision structures expression of Equate-to-differentiate theory (Li, 2004).

**(iii). Second Order Substitution, $\tau''(G)$:**

Section 1. Options Paralleling

$$G_v = \begin{pmatrix} a_i' a_j' & 1 \\ 1 & 1 \end{pmatrix} \quad (39)$$

$$G_v = \begin{pmatrix} 1 & b_i' b_j' \\ 1 & 1 \end{pmatrix} \quad (40)$$

$$G_v = \begin{pmatrix} 1 & 1 \\ 1 & c_i' c_j' \end{pmatrix} \quad (41)$$

Up to now we considered the structure of decision matrix $G$.

$$(12) \circ (39) = \begin{pmatrix} a_i a_j a_i' a_j' & 1 \\ 1 & 1 \end{pmatrix} \quad (42)$$

$$(12) \circ (40) = \begin{pmatrix} a_i a_j & b_i' b_j' \\ 1 & 1 \end{pmatrix} \quad (43)$$

$$(12) \circ (41) = \begin{pmatrix} a_i a_j & 1 \\ 1 & c_i' c_j' \end{pmatrix} \quad (44)$$

$$(13)\ °(39) = \begin{pmatrix} a_i'a_j' & b_ib_j \\ 1 & 1 \end{pmatrix} \quad (45)$$

$$(13)\ °(40) = \begin{pmatrix} 1 & b_ib_jb_i'b_j' \\ 1 & 1 \end{pmatrix} \quad (46)$$

$$(13)\ °(41) = \begin{pmatrix} 1 & b_ib_j \\ 1 & c_i'c_j' \end{pmatrix} \quad (47)$$

$$(14)\ °(39) = \begin{pmatrix} a_i'a_j' & 1 \\ 1 & c_ic_j \end{pmatrix} \quad (48)$$

$$(14)\ °(40) = \begin{pmatrix} 1 & b_i'b_j' \\ 1 & c_ic_j \end{pmatrix} \quad (49)$$

$$(14)\ °(41) = \begin{pmatrix} 1 & 1 \\ 1 & c_ic_jc_i'c_j' \end{pmatrix} \quad (50)$$

In this case, the probability matrix and value matrix have same position substitution. Thus, we cannot compare the A (Pi, Vi) and B (Pj, Vj) immediately. From assumptions 1, therefore, we have to substitute probability or value matrix further, during decision making. Let us considered (42) that if Vj > Vi, $a_j'$ will be substituted for $b_j'$. Then, (42) can be described as (18), $\tau$ ($G_{18}$) =$\tau$' ($G_{42}$) $d\tau$. In the same way, all of structure $\tau$'' (G) can be deduced to structure $\tau$' (G).

Section 2. Options Crossing

$$G_v = \begin{pmatrix} a_j' & b_i' \\ 1 & 1 \end{pmatrix} \quad (51)$$

$$G_v = \begin{pmatrix} a_j' & 1 \\ 1 & c_i' \end{pmatrix} \quad (52)$$

$$G_v = \begin{pmatrix} 1 & b_j' \\ 1 & c_i' \end{pmatrix} \quad (53)$$

Up to now we considered the structure of decision matrix G.

$$(5)\ °(51) = \begin{pmatrix} a_ia_j' & b_jb_i' \\ 1 & 1 \end{pmatrix} \quad (54)$$

$$(5)\ °(52) = \begin{pmatrix} a_ia_j' & b_j \\ 1 & c_i' \end{pmatrix} \quad (55)$$

$$(5)\ °(53) = \begin{pmatrix} a_i & b_jb_j' \\ 1 & c_i' \end{pmatrix} \quad (56)$$

$$(6)\ °(51) = \begin{pmatrix} a_ia_j' & b_i' \\ 1 & c_j \end{pmatrix} \quad (57)$$

$$(6)\ °(52) = \begin{pmatrix} a_ia_j' & 1 \\ 1 & c_i'c_j \end{pmatrix} \quad (58)$$

$$(6)\ °(53) = \begin{pmatrix} a_i & b_j{'} \\ 1 & c_j c_i{'} \end{pmatrix} \quad (59)$$

$$(7)\ °(51) = \begin{pmatrix} a_j{'} & b_i b_i{'} \\ 1 & c_j \end{pmatrix} \quad (60)$$

$$(7)\ °(52) = \begin{pmatrix} a_j{'} & b_i \\ 1 & c_i{'} c_j \end{pmatrix} \quad (61)$$

$$(7)\ °(53) = \begin{pmatrix} 1 & b_i b_j{'} \\ 1 & c_j c_i{'} \end{pmatrix} \quad (62)$$

Unfortunately, in this case, we could not immediately compare the A (Pi, Vi) and B (Pj, Vj). In this structure, there are two sub-structures, such as options crossing, options paralleling. Actually, this structure leads to most of non-compensatory theories instability.

We have to acknowledge that the most efficient way to reduce the order of decision structure is calculating expected value. Sometimes, we make decision based on a compensatory expectation-maximizing process. That is, we will calculate $EV_i$= $Pi \times Vi$ and $EV_j$= $Pj \times Vj$. After that, we will compare $EV_i$ with $EV_j$, and then choose the bigger one. However, the calculate process, which is expending vast amounts of energy. It is a dilemma problem if $EV_i = EV_j$ or we are not able to compare relationship between $EV_i$ and $EV_j$ immediately. More precisely, decision making is an order reduction process, which is simplifying the decision structure. Thus, the deliberation time (Busemeyer & Townsend, 1993) at least has four parts, which are consist of substitution time $t_0$, $\tau"$ (G) $d\tau$ time $t_1$, $\tau'$ (G) $d\tau$ time $t_2$ and calculation time $t_{cal}$.

In addition, preference reversal is an unexplained phenomenon. The preference reversal often happens on the second order substitution, options crossing or paralleling. At this time, we have to reduce the order of decision structure. For instance, A (Pi, Vi) and B (Pj, Vj),

$$\tau"(G_{54}) = \begin{pmatrix} a_i a_j{'} & b_j b_i{'} \\ 1 & 1 \end{pmatrix} \quad (63)$$

$$\overset{d\tau}{\Rightarrow} \begin{cases} \tau'(G18) = \begin{pmatrix} a_i a_j a_j{'} & b_i{'} \\ 1 & 1 \end{pmatrix} \\ \tau'(G21) = \begin{pmatrix} a_j{'} & b_i b_j b_i{'} \\ 1 & 1 \end{pmatrix} \\ \tau'(G30) = \begin{pmatrix} a_i a_j{'} a_i{'} & b_j \\ 1 & 1 \end{pmatrix} \\ \tau'(G31) = \begin{pmatrix} a_i & b_j b_i{'} b_j{'} \\ 1 & 1 \end{pmatrix} \end{cases} \quad (64)$$

$$\overset{d\tau}{\Rightarrow} \begin{cases} \tau(G) = \begin{pmatrix} a_i a_i{'} & b_j b_j{'} \\ 1 & 1 \end{pmatrix} \rightarrow B > A \\ \tau(G) = \begin{pmatrix} a_j a_j{'} & b_i b_i{'} \\ 1 & 1 \end{pmatrix} \rightarrow A > B \end{cases} \quad (65)$$

Let us consider second order substitution (63) that the decision structure changing from second order to zero order. The different results were deduced from different paths. We are very certainly that the order reduction is an implicit process. Namely, we are not really sure which path will occur during decision making process. It is why preference reversal always happens when making decisions.

From the above assumption, we do not want to discuss merits or drawbacks nor and just want to explain most of current theory. We have now deduced the requisite laws of the decision structure, and we proceed to show application to compensatory theories of risky choice

### III. STRUCTURE DECISION OF COMPENSATORY

Mainstream theories of decision making under risk hold that risky choices are based on a compensatory expectation-maximizing process. Furthermore, let us rethink these compensatory theories which reveal the relationship between options, $S_p$ and $S_v$, and substitution elements, $G_p$ and $G_v$.

According to structure of decision matrix, there are two options:

$$A (P_i, V_i) \text{ vs. } B (P_j, V_j)$$

Let us assume that

$$a = k_1 P_i = k_1' P_j \quad (66)$$
$$b = k_2 P_i = k_2' P_j \quad (67)$$
$$c = k_3 P_i = k_3' P_j \quad (68)$$
$$a' = m_1 V_i = m_1' V_j \quad (69)$$
$$b' = m_2 V_i = m_2' V_j \quad (70)$$
$$c' = m_3 V_i = m_3' V_j \quad (71)$$

As we know, a new one will be substituted for an appropriate element of the decision matrix, and from (66) - (71), it immediately follows

$$aa' = k_1 m_1 P_{i_a} V_{i_a} = k_1' m_1' P_{j_a} V_{j_a} \quad (72)$$
$$bb' = k_2 m_2 P_{i_b} V_{i_b} = k_2' m_2' P_{j_b} V_{j_b} \quad (73)$$
$$cc' = k_3 m_3 P_{i_c} V_{i_c} = k_3' m_3' P_{j_c} V_{j_c} \quad (74)$$

**Corollary 1**

Let us discuss particular types of above equations:
if $k = k' = m = m' = 1$,

$$aa' = P_{i_a} V_{i_a} = P_{j_a} V_{j_a}$$
$$bb' = P_{i_b} V_{i_b} = P_{j_b} V_{j_b}$$
$$cc' = P_{i_c} V_{i_c} = P_{j_c} V_{j_c}$$

In this case, it is the decision structures expression of the EV.

**Corollary 2**

if $k = k' = 1$ and $m$ & $m' \neq 1$,

$$aa' = m_1 P_{i_a} V_{i_a} = m_1' P_{j_a} V_{j_a}$$
$$bb' = m_2 P_{i_b} V_{i_b} = m_2' P_{j_b} V_{j_b}$$
$$cc' = m_3 P_{i_c} V_{i_c} = m_3' P_{j_c} V_{j_c}$$

In this case, it is the decision structures expression of the Expected Utility (von Neumann & Morgenstern, 1947).

**Corollary 3**

if k & k' ≠ 1 and m & m' ≠ 1,
$$aa' = k_1 m_1 Pi_a Vi_a = k_1'm_1'Pj_a Vj_a$$
$$bb' = k_2 m_2 Pi_b Vi_b = k_2'm_2'Pj_b Vj_b$$
$$cc' = k_3 m_3 Pi_c Vi_c = k_3'm_3'Pj_c Vj_c$$

In this case, it is the decision structures expression of the Prospect Theory (Kahneman & Tversky, 1979).

## IV. APPLICATION TO MORE DECISION PHENOMENA

### IV. A. Allais Paradox

Firstly, the Allais paradox consists of two pairs of choices, each pair having two alternative prospects (Allais, 1953). The first pair of Allais paradox is, A (Pi = 1, Vi = 1 000 000) vs. B ($Pj_1$ = .1, $Vj_1$ = 5 000 000; $Pj_2$ = .89, $Vj_2$ = 1 000 000; $Pj_3$ = .01, $Vj_3$ = 0).

Let us consider the structure of Allais paradox,
$$Pi \rightarrow 1, Vi \rightarrow c_i';$$
$$Pj_1 \rightarrow a_j, Vj_1 \rightarrow c_j';$$
$$Pj_2 \rightarrow b_j, Vj_2 \rightarrow c_j';$$
$$Pj_3 \rightarrow a_j, Vj_3 \rightarrow 0.$$

Thus,
$$G_I = \begin{pmatrix} a_j a_j & b_j \\ 1_i & c_i'c_j'c_j' \end{pmatrix} \quad (75)$$

In this case, we can deduce, the option A always has superiority position in the decision matrix, that the option A > B.

The second pair of choice is, C ($Pi_1$ = .11, $Vi_1$ = 1 000 000; $Pi_2$ = .89, $Vi_2$ = 0) vs. D ($Pj_1$ = .1, $Vj_1$ = 5 000 000; $Pj_2$ = .9, $Vj_2$ = 0).

Where,
$$Pi_1 \rightarrow a_i, Vi_1 \rightarrow c_i';$$
$$Pi_2 \rightarrow c_i, Vi_2 \rightarrow 0;$$
$$Pj_1 \rightarrow a_j, Vj_1 \rightarrow c_j';$$
$$Pj_2 \rightarrow c_j, Vj_2 \rightarrow 0;$$

Thus,
$$G_{II} = \begin{pmatrix} a_i a_j & 1 \\ 1 & c_i c_j c_i' c_j' \end{pmatrix} \quad (76)$$

In this case, the (76) also can be described as (44). Namely, the (76) is a variation structure of the two order substitution. As mentioned above, all of two order substitution structures can be reduced to one order substitution structures.

$$G_{II} = \begin{pmatrix} a_i a_j & 1 \\ 1 & c_i c_j c_i' c_j' \end{pmatrix} \xrightarrow{d\tau} \begin{pmatrix} a_i a_j & 1 \\ 1 & c_i' c_j' \end{pmatrix} \xrightarrow{d\tau} \begin{pmatrix} a_i a_j & b_i' \\ 1 & c_j' \end{pmatrix} \quad (77)$$

Thus, we can deduce that option C < D.

### IV. B. The Performance of π Function

As we know, Kahneman and Tversky (1979) was discussed the weighting function π, which relates decision weights to stated probabilities. They believe that large and intermediate probabilities are underweighted while small probabilities are overweighted.

Let us assume that the elements of probability matrix consist of $\{0, a = .3, b = .5, c = .7, 1\}$. In addition, Pi is a small probability, from 0 to 0.35; Pj is an intermediate probability, from .35 to .7; Pk is a large probability, from .7 to 1.

So, let us discuss the substitution of probability matrix.

Where,

$$G_p = \begin{pmatrix} a & b \\ 1 & c \end{pmatrix}, \, a < b < c < 1.$$

$$Pi \rightarrow a = .3;$$
$$Pj \rightarrow b = .5;$$
$$Pk \rightarrow c = .7.$$

Thus,
(a) If $Pi < a$, the small probabilities are overweighted.
(b) If $Pj < b$, the intermediate probabilities are overweighted.
(c) If $Pj > b$, the intermediate probabilities are underweighted.
(d) If $Pk > c$, the large probabilities are underweighted.

From the above assumption, we can deduce why the weighting function π has these performance. However, it is known that the value range is larger than probability range in experiments. Obviously, the substitutive range is, $a' < b' < c'$. That is, we can also deduce, from value matrix, that the value function is generally concave for gain. In addition, the decision structure can simply explain the phenomenon of subadditive, subcertainty and subproportionality.

### IV. C. Ellsberg Paradox

Ellsberg (1961) propose that suppose you have an urn containing 30 red balls and 60 other balls that are either black or yellow. The balls are well mixed so that each ball is as likely to be drawn as any other.

Gamble A: You receive $100 if you draw a red ball;
Gamble B: You receive $100 if you draw a black ball;
Gamble C: You receive $100 if you draw a red or yellow ball;
Gamble D: You receive $100 if you draw a black or yellow ball.

Let us consider the structure of Ellsberg paradox,

$$\text{Gamble A:} \quad P_A = \frac{1}{3}, V_A = 100$$

$$\text{Gamble B:} \quad P_B = p, V_B = 100, P_B \in (0, \frac{2}{3})$$

$$\text{Gamble C:} \quad P_C = q, \ V_C = 100, \ P_C \in (\tfrac{1}{3}, 1)$$

$$\text{Gamble D:} \quad P_D = \tfrac{2}{3}, \ V_D = 100$$

Then, let us consider the substitution of decision matrix.

$$P_A = \tfrac{1}{3} \ \to \ a_A$$

$$P_B \ \to \ a_B \text{ or } b_B, \ P_B \in (0, \tfrac{2}{3})$$

$$P_C \ \to \ b_C \text{ or } c_C, \ P_C \in (\tfrac{1}{3}, 1)$$

$$P_D = \tfrac{2}{3} \ \to \ b_D$$

$$V \ \to \ b'$$

Thus,

$$G_{AB} = \begin{pmatrix} a_A a_B & b_A' b_B' \\ 1 & 1 \end{pmatrix} \quad (78)$$

$$G_{AB'} = \begin{pmatrix} a_A & b_B b_A' b_B' \\ 1 & 1 \end{pmatrix} \quad (79)$$

$$G_{CD} = \begin{pmatrix} 1 & b_C b_D b_C' b_D' \\ 1 & 1 \end{pmatrix} \quad (80)$$

$$G_{CD'} = \begin{pmatrix} 1 & b_D b_C' b_D' \\ 1 & c_C \end{pmatrix} \quad (81)$$

Ellsberg's results have been replicated many times, people strictly prefer Gamble A to B, and prefer Gamble D to C. The most of researchers believe that the experimental evidence suggests that people prefer risk to uncertainty. That is, according to decision structure that the result of (78), equivalent to (53), is A > B, and result of (80), equivalent to (46), is D > C. Therefore, the experimental evidence suggests that

$$P_B \ \to \ a_B, \ P_B < P_A = \tfrac{1}{3} \ ;$$

$$P_C \ \to \ b_C, \ P_C < P_D = \tfrac{2}{3} \ ;$$

$$P_{\text{yellow}} \ \to \ a_{yellow}, \ P_{\text{yellow}} < P_A = \tfrac{1}{3}.$$

Namely that, each probability of ball, draw a black ball (Gamble B) or draw a yellow ball (Gamble C), is an individual event.

Thus,

$$P_B = p = a_B$$

$$P_C = q = \tfrac{1}{3} + a_C \neq 1 - p$$

From the above assumption, we can deduce two requisite laws of the decision group that

a) A new probability will be substituted for an appropriate element of probability matrix, when we facing a risky situation.
b) A new uncertainty probability will be substitute for a minimal element of

probability matrix, when we facing an uncertainty situation.

## V. CONCLUSION

It is understandable that the better strategy is conservatively estimated the possibility of gain when we facing uncertainty from an evolutionary perspective. Actually, the Ellsberg paradox is not a real paradox, but a clue to understand what risk and uncertainty are. In the next step, we have to predict that the substitution, structure order reduction and calculation process might act on specific regions of the brain to regulate decision making. But the substitution and order reduction may not be separated out and directly observe with current technology. We have to discuss people makes decisions based on number or concept or both in the future research.

Finally, life is complex and dynamic, we are almost certainly that the above assumption just discuss from a particular and simplest condition, which is a second order matrix. Along with the theory development, we have to reconsider what risky choice and decision structure are. We can deeply understand this complex process if all decision structures can be perfectly known. Our experience is far from being sufficient to enable us to accurately explanation what the decision making is. But, we wish to provide a different with the former method.

In course of time, the mathematical consequences will be gradually deduced, and decision structure concepts, will be reconciled to the new ideas of decision making, in the prospect that they may lead to pre-established harmony between mathematics and decision making.